# EVOLUTIONARY ALGORITHMS IN GENETIC REGULATORY NETWORKS MODEL

**Khalid Raza*, Rafat Parveen**
Department of Computer Science, Jamia Millia Islamia (Central University), New Delhi, India
***Corresponding author:** Email: kraza@jmi.ac.in

**ABSTRACT:**

Genetic Regulatory Networks (GRNs) plays a vital role in the understanding of complex biological processes. Modeling GRNs is significantly important in order to reveal fundamental cellular processes, examine gene functions and understanding their complex relationships. Understanding the interactions between genes gives rise to develop better method for drug discovery and diagnosis of the disease since many diseases are characterized by abnormal behaviour of the genes. In this paper we have reviewed various evolutionary algorithms-based approach for modeling GRNs and discussed various opportunities and challenges.

## INTRODUCTION

With the advancement in DNA microarray technologies, it has become possible to effectively and efficiently measure gene expression levels of up to tens of thousands of genes under various conditions simultaneously; which are playing successfully role in gene function prediction, drug development, disease diagnosis and patient survival analysis [4][5]. Systems biology has grown up as new discipline which combines the research efforts in biology, chemistry, physics, mathematics, computer science, and other disciplines to systematically study the surprisingly complex interactions in biological systems. In the last decade, many efforts have been applied in order to develop computational methods for inferring underlying genetic networks from time series gene expression data. Today inference of genetic networks (also called reverse-engineering of genetic networks) from time series gene expression data has become rather important in order to understand the complex biological interactions and behaviour. Many theoretical models have been proposed to model, analyze, reconstruct and infer complex regulatory interactions and provide hypothesis for experimental verification. These models are Boolean networks, differential equations, Bayesian networks, Petri nets, machine learning approaches, evolutionary computing etc. Most of the interactions and parameters of these networks are still not well-known or poorly known, so accurately inferring such networks remains a challenge for the researchers.

The motivation behind the study of genetic network is that it describes how genes or groups of genes interact with each other and to identify the complex regulatory interaction between genes in a living organism. The genetic networks enable us to understand the intricate interactions of multiple genes under various stimuli or environmental conditions [1]. Inference of genetic network is important because it presents a synthetic network view of the current biological knowledge and allows structuring in such a way that reveals relevant properties that might remain hidden otherwise. Genetic network also allows prediction of dynamic behaviour of the network which is later compared with experimental results and allows either confirmation of the model's accuracy or recommend correction in the model. System Biology is a fast growing interdisciplinary research area which tries to decipher the complex interactions between cell products [2].

A large number of mathematical modeling techniques and inferential algorithm have been devised. Basic steps for modeling genetic regulatory networks (GRNs) consists of few main steps such as i) choosing an appropriate model ii) inferring parameters from the data iii) validating the model iv) conducting simulations of the GRN to predict its behaviour under various conditions [3]. For modeling GRN, genes are treated as variables which change their expression values with respect to time. The rest of the paper is organized as follows. Section II discusses some of the basic modeling techniques, such as Boolean networks, generalized Bayesian networks, linear and non-linear differential equations, Petri net, fuzz logic, artificial neural networks etc. Section III gives a brief concept of about evolutionary algorithms. Section IV discusses the role of evolutionary algorithms (EA) and its hybridization in gene regulatory networks modeling. Finally Section V concludes the paper by comparing various EA-based approaches, their merits and flaws and highlighting some of the challenges and future trends.

[II] MODELING TECHNIQUES

There are several approaches for modeling and inferring GRNs from gene expression data including directed graphs model, Boolean networks [6]–[8], generalized Bayesian networks [9][10], linear and non-linear ordinary differential equations (ODEs) [11]–[15], Petri net [16][17], fuzzy logic, artificial neural networks etc.

*Directed graph* is the most simple and straightforward way to model a GRN. Here vertices of the directed graph represent genes and edges denote interactions among the genes. A directed edge is defined as a tuple $(i, j, s)$, where $i$ denotes the head, $j$ tail of the edge and $s$ is equal to either + or – indicating whether $i$ is activated or inhibited by $j$. The graph representation of GRNs allows a number of operations that can be carried out to make prediction about biological processes [12]. Figure 1 shows a directed graph representation of GRNs.

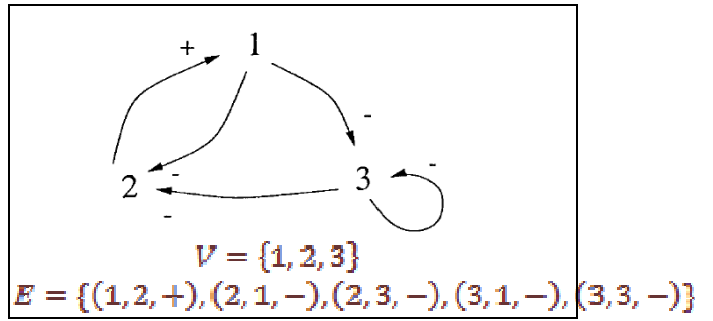

**Figure 1.** Directed graph representing a GRN and its definition

*Boolean networks* are based on Boolean logic where each gene is either fully expressed (represented as "1") or not expressed at all (represented as "0"). The interactions between elements can be represented by Boolean functions which calculate the state of a gene from the activation of other genes. Hence, network can transition from one state to another state. The advantages of Boolean network models are its simplicity and finite state space but unfortunately these models are unable to capture the effect of genes at intermediate levels. It also unrealistically assumes that transitions between activation states of the genes are synchronous. In Boolean networks, if there are N genes in a GRN, the network can be in any of the $2^N$ possible states. This exponential state space makes the classification of attractors a computationally and memory intensive task. Figure 2 shows a Boolean network of three entities a, b and c. The state transitions follow the regulation functions shown on the right, which describe the rules of the model. Thin arrows indicate the regulators of each node and time steps are represented by thick arrows [19].

*Bayesian networks* (BNs) models use directed acyclic graph $G = (V, E)$ to represent the network where vertices ($V$) corresponds to genes and the edges ($E$) presenting the conditionally dependent interactions between genes. These models estimate the multivariate joint probability distributions through local probabilities [18]. The main advantages of these models are its capabilities to deal with the stochastic aspects of gene expression, handling noisy and incomplete data. However, these models are unable to deal with the dynamic aspects of gene regulations. To overcome this dynamicity problem Dynamic Bayesian networks (DBNs) were devised.

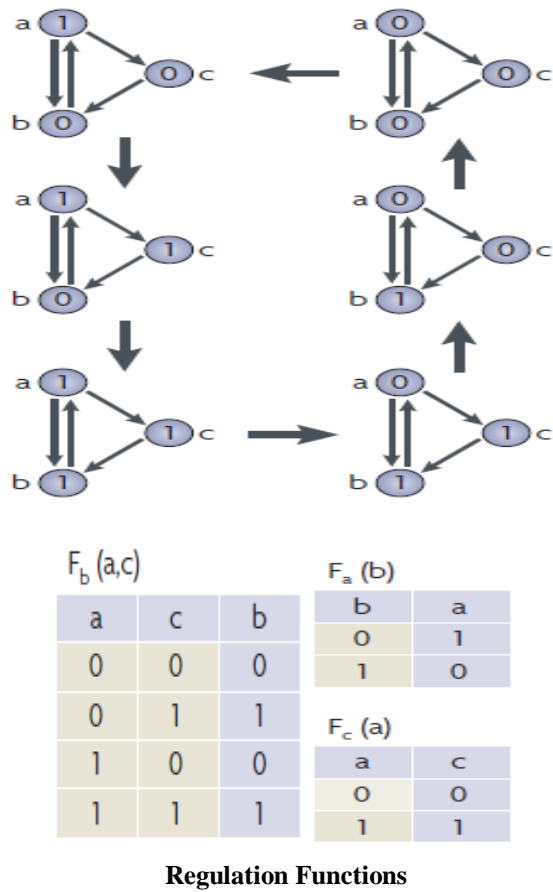

**Regulation Functions**

**Figure 2.** A Boolean network with regulation functions of three elements a, b and c [19]

*Ordinary differential equations* (ODEs) have been most widely used techniques for modeling dynamic biochemical systems [18] especially genetic regulatory networks [12]. The ODEs allow more detailed descriptions of network dynamics, by explicitly modeling the concentration changes of molecules over time. The ODE formalism models the concentrations of RNAs, proteins, and other molecules by time-dependent variables. Regulatory interactions take the form of functional and differential relations between the concentration variables. Although ODE approach provides detailed information about the network's dynamics but it requires high-quality data on kinetic parameters and hence it is currently applicable to only a few systems. A detailed discussion and review about various differential equation-based models can be found in [12] and [19].

*Petri net* is a non-deterministic approach for modeling the dynamics of regulatory networks. Petri nets are an extension to graph models that represents a well-established technique for modeling regulatory systems. Following the analogy of biological systems, Petri nets have successfully been applied for simulating gene regulatory network, allowing simple quantitative representation of dynamic processes. The drawback of Petri nets model is that it does not support hierarchical structuring, which makes them difficult to use for large-scale models.

*Fuzzy logic* is one of the constituents of soft computing technique which has been drawn from engineering and other applied sciences. Fuzzy logic is suitable for modeling GRN because i) fuzzy logic extracts trends not values so fuzzy logic is inherently tolerant to noisy data ii) it is computationally efficient and can be extended to large number of components and iii) it is implemented in a user-friendly language (e.g. if-then rules). Fuzzy logic has been successfully used for modeling gene regulatory networks due to its capability to represent non-linear systems, its friendly language to incorporate and edit domain knowledge in the form of fuzzy rules [20]–[24]. However, computational time is a major obstacle in developing more complex fuzzy models. One solution to this problem can be preprocessing of data so that computation time can be reduced. Most widely used method of data pre-processing is data clustering. A wide range of clustering algorithms bas been proposed in the literature including hierarchical clustering, adaptive resonance theory (ART) [51], self-organizing maps (SOM), k-means, and fuzzy c-means.

*Artificial neural networks* (ANNs) have been developed as generalization of mathematical models of biological nervous systems. The capabilities of ANNs to learn from the data-rich environment, approximate any multivariate nonlinear function and its robustness to noisy data make it a suitable choice for modeling GRNs from gene expression data. Neural network architecture has a number of nodes and wirings between them. Generally, the number of nodes is defined as the number of genes but it may also represent any other factors involved in the regulatory network. Let a N-dimensional vector $u(t)$ be the expression state of a gene network containing N genes and element $u_j(t)$ is the expression state of gene $j$ at time $t$. The wirings define regulatory interactions between genes, which are represented by a weight matrix $w$. A wiring from gene $j$ to gene $i$ means a non-zero weight $w_{ij}$. A positive weight implies a stimulating effect (positive feedback) while a negative weight implies repression (negative feedback). A zero weight $w_i$; means no regulatory interaction. The control strength is the multiplication of weight $w_{ij}$ and state value $u_j$. The total regulatory input to gene $i$ is the sum of regulatory strengths of all genes in the regulatory network [52],

$$r_i(t) = \sum_{j=1}^{N} w_{ij} u_j(t) + \alpha_i, \quad i = 1, 2, ..., N$$

where $\alpha_i$ is a parameter to represent the influence of external inputs or reaction delay. A squashing function then transfers the regulatory input $r(t)$ into a normalized transcriptional response. Various types of ANNs have been successfully applied for modeling gene regulatory interactions including perceptrons [25]–[27], self-organizing maps [28]–[29] and recurrent neural networks (RNNs) [30][31].

**BASICS OF EVOLUTIONARY ALGORITHMS**

*Evolutionary algorithms* (EAs) are optimization algorithm based on Darwin's theory of evolution. The field of evolutionary algorithms has been growing rapidly over the last few years. It is basically a search algorithm that is modeled on the mechanics of natural selection and natural genetics. It combines survival of the fittest among individuals with a structured yet randomized information exchange to form a search algorithm. In EAs optimization techniques searching from a population are done from a single point and for each iterations a competitive selection is done. The solutions with high "fitness" are recombined with other solutions. The solutions are then "mutated" by making a small change to a single element of the solution. The main purpose of recombination and mutation is to generate new solutions but it is biased towards regions of the space for which good solutions have already been identified. Generally, three evolutionary techniques are distinguished: genetic programming (GP), genetic algorithms (GA) and evolutionary programming (EP) (Goldberg, 1989, Michalewicz, 1996). Genetic programming focuses on programs evolution, genetic algorithms focuses on optimizing general combinatorial problems and evolutionary programming focuses on optimizing continuous functions without recombination. EAs belong to the class of probabilistic algorithms and they differ from random algorithms in that they combine elements of directed and stochastic search. Due to this reason EAs are more robust than directed search methods. Another merit of EAs is that they maintain a population of potential solutions while other search techniques process a single point of the search space. The limitation of GP and GA-based modeling techniques are that they do not take care of the noise effect which is quite common in microarray data. Figure 3 shows a general scheme of EAs and detailed discussion on EAs can be found in [32][33]. Various constituents of EAs have been successfully applied for modeling GRNs [34]–[51].

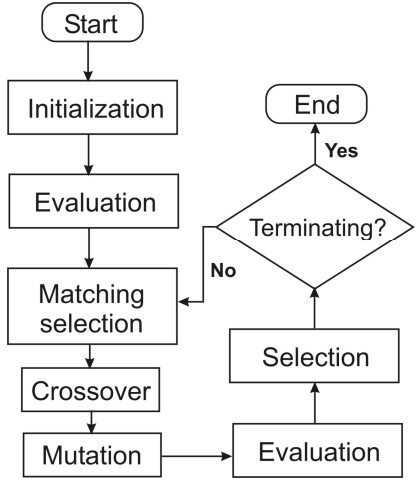

**Figure 3. A general scheme of evolutionary algorithm**

## ROLE OF EVOLUTIONARY ALGORITHMS IN GRN MODELING

Several constituents of evolutionary algorithms such as genetic programming, genetic algorithm, evolutionary programming and their hybridization have been successfully applied for efficient modeling genetic regulatory networks. A combination of Genetic Programming (GP) and Least Mean Square (LMS) method, called LMS-GP, to identify a concise form of regulation between genes from time series data has been applied by Ando *et.al.* [35]. The LMS methods are applied to determine the coefficients of the GPs, which decreases the Mean Squared Error (MSE) between the observed and model time series without complicating the GPs. The proposed LMS-GP model has been tested on artificial as well as real-world data. The model has an average MSE of $4.21 \times 10^{-3}$ over 10 runs, while standard GP averaged MSE is $6.704 \times 10^{-3}$ over 10 runs. Genetic programming jointly with Kalman filtering approach to infer GRNs from time series data has been applied by Wang *et.al.* [36]. In this method, nonlinear differential equation model is adopted and an iterative algorithm has been proposed to identify the model. Here GP is employed to identify the structure of the model and Kalman filtering is deployed to estimate the parameters in each of the iterations. The proposed model has been tested using synthetic as well as time-series gene-expression data of yeast protein synthesis. Due to availability of noise in microarray data, the Kalman filtering may not be appropriate for estimating parameters.

Decoupled S-system formalism for inferring effective kinetic parameters from time series data and Trigonometric Differential Evolution (TDE) as the optimization engine for capturing the dynamics of gene expression data has been applied by Noman *et.al.* [37]. Here fitness function used is a modified version of Kimaru *et.al.* [38] for minimizing the number of false positive predictions. Spare network structure has been identified with the help of hill-climbing local search (HCLS) method within the framework of proposed EA. Experiments on small scale artificial network in noise-free as well as noisy environment is done and found that proposed model successfully identify the network structure and its parameter values. Real-life data has also been used for reconstructing the SOS DNA repair network of *E.coli*. The proposed model correctly identified the regulations of gene lexA and some other known regulations. The doubled S-system model proposed by Chowdhury and Chetty [45] extended the work of Noman *et.al.* [37]. In this model, GA is used for scoring the networks' several useful features for accurate inference of network, such as a Prediction Initialization (PI) algorithm to initialize the individuals, a Flip Operation (FO) for matching the values, and a restricted execution of HCLS over few individuals. It also proposes a refinement technique for optimizing sensitivity and specificity of inferred networks [45].

Maeshiro *et.al.* [34] proposed an EAs based approach to predict GRNs from gene expression time course data which consists of two stage loop each using following steps: i) generate M initial fictitious networks ii) simulate each network with ultra high speed simulator Starpack [47] (Signal Transduction Advanced Research Package) and find out the differences from biological experimental data iii) finish, if reaches maximum number of loops or the score of the top ranked network shows no improvement, iv) execute roulette selection, mutate and execute crossover to generate a new set of networks. Here simulation of the second stage has higher precision over first stage, serving as local optimization process. The proposed model was tested on five synthetic networks and results were compared with dynamic Bayesian network model and found that sensitivity is approximately 5% higher and precision was approximately equal. The prediction of four networks resulted in the sensitivity of 70–80%, and precision of 70%.

Ho *et.al.* [40] proposed an intelligent two-stage evolutionary algorithm (iTEA) to infer the S-system models of N-gene genetic networks from time-series data of gene expression. The problem is initially decomposed into N optimization subproblems having 2(N+1) parameters each. In the first stage, each subproblem is solved using EA and intelligent crossover based on an orthogonal arrays and factor analysis. In the second stage, solutions of the N subproblems are combined and refined using an OED-based simulated annealing framework for handling noisy gene expression data. The efficiency of iTEA was evaluated using simulated expression patterns with and without noise. Huang *et.al.* [49] extended the method iTEA proposed by [40] by generating additional multiple data sets of gene expression profiles by perturbing the given data set and named it as iTEAP method. The iTEAP model copes up "multiplicity of solutions" by using noisy duplicates to obtain accurate and robust GRNs. Shu *et.al.* [46] has introduced an improved version of iTEA method named it as iTEA2 for establishing large-scale GRN by incorporating gene regulation domain knowledge in EA. The iTEA2 method uses hybridize encoding scheme that comprises of regulation strength, gene number regulated, and binary control parameters in a chromosome where value of the strength parameter is the kinetic order and control parameters indicate that the up- and down-regulated kinetic orders are active or not. If I be the maximal number of genes that directly regulated each gene then total number of parameters encoded in a chromosome for one gene is 5I+2, which is independent of N. Hence, the number of parameters needed for the construction of GRN reduced from $O(N^2)$ to $O(N)$ [46].

Chan *et.al.* [41] extracted GRNs from time-series gene expression data using a two-stage methodology that was implemented in the software tool "Gene Network Explorer (GNetXP)". At the first stage, GA has been applied in selecting the initial cluster centers for subsequent Expectation Maximization (EM) partitioning. At the second stage, Kalman Filter was deployed to identify a set of first-order differential equations which describes the dynamics of the network and used these equations for determining important gene interactions and predicting gene expression values at future time points. The proposed methodology was tested on the human fibroblast response gene expression data.

Tominaga and Horton [42] developed algorithm for inferring biological networks using only time-series data restricted to a scale-free networks. The S-System was adopted as network model and distributed GA to optimize the model. The inherent parallelism enhances the performance of the model [18]. Ram and Chetty [43] proposed a causal GA-based approach for learning GRNs which is guided by exploiting certain characteristics of diversity and heuristic in order to generate better networks. The proposed formalism is named as guided genetic algorithm (GGA) approach.

describe their regulation type. In this hybridized approach, recurrent network, self-organizing structure and evolutionary training generate an optimum pool of regulatory relationships and fuzzy systems tolerate noise-related issues.

One of the main objectives of understanding gene regulation is to reveal how combination of transcription factors (TFs) control sets of co-expressed genes under specific experimental conditions. Schroder et.al. [50] proposed multi-objective genetic algorithms (MOGAs) for

| Techniques applied | Results obtained | References |
|---|---|---|
| Multi-objective GA + biclustering | Extracted GRN using correlation | Mitra et.al., 2009 [39] |
| Multi-objective GA (MOGA) | Infers transcriptional regulators for sets of co-expressed genes | Schröder et.al. 2011 [50] |
| Distributed GA (named as iTEA) | Divided problem into N optimization subproblems | Ho et.al. 2007 [40] |
| EA + domain knowledge (named as iTEA2) | Infers large-scale GRNs | Shu et.al. 2011 [46] |
| GA + clustering | Modeled GRN using Kalman filters | Chan et.al. 2006 [41] |
| GA+ Fuzzy clustering | Reconstructs GRNs | Shoaib et.al. 2006 [44] |
| Guided GA (GGA) | Discovered GRNs | Ram & Chetty, 2007 [43] |
| Multilayer ENFRN | Captures potential regulators of target genes and describe their regulation type | Maraziotis et.al. 2010 [48] |
| GP + LMS | Describes a causal model for GRNs | Ando et.al. 2002 [35] |
| GP + Kalman Filtering | Infers GRNs | Wang et.al. 2006 [36] |
| Doubled S-System + TDE | Captures the dynamics of GRNs | Noman et.al. 2005 [37] |
| Distributed GA + S-system + domain knowledge | Modeled GRN in a scale-free network | Tominaga & Horton, 2006 [42] |
| Double S-System + PI + FO | Reconstructs GRNs | Chowdhury & Chetty 2011 [45] |

Table: 1. Hybridized form of GAs for Modeling GRNs

Mitra et.al. [39] proposed multiobjective evolutionary biclustering and correlation-based approach to extract gene interaction networks from microarray data. Biclustering has been applied to find a subset of similarly expressed genes under some specific experimental condition. To add/remove relevant/irrelevant genes for fine tuning, local search strategy has been deployed. Preprocessing technique is applied to preserve strongly correlated gene interaction pairs.

A multilayer evolutionary trained neuro-fuzzy recurrent network (ENFRN) approach was proposed by Maraziotis et.al. [48] which captures potential regulators of target genes and inferring transcriptional regulators for sets of co-expressed genes. In this work, three objective functions have been designed for stimulus response and can be used to integrate a priori knowledge into the detection of gene regulatory modules. The proposed method was tested and evaluated on whole genome microarray measurements of drug-response in human hepatocytes [50].

Adaptive Fuzzy Evolutionary GRN Reconstruction (AFEGRN) framework has been developed by Shoaib et.al. [44] for modeling GRNs. The AFEGRN framework is able to automatically determine model parameters, for example, number of clusters for fuzzy c-means using fuzzy-PBM index and estimation of Gaussian Distribution Algorithm. The proposed model was tested using breast cancer data to demonstrate effectiveness of AFEGRN to model any GRN. The proposed framework composed of six steps: i) data preprocessing ii) computation of number of clusters iii) clustering using fuzzy c-means iv) gene selection v) GRN construction and vi) GRNs comparison. Table 1 shows a comparative view of various hybridized form of GA-based approach and their results obtained. EA-based approaches are generally hybridized, with several other techniques such as clustering, least mean square, Kalman filtering, Trigonometric Differential Evolution, etc, for better results.

## CONCLUSIONS & CHALLENGES

The models of gene regulatory networks are developed to capture the behavior of the system being modeled, and it is also able to produce predictions corresponding with experimental observations. Understanding GRNs is essential because a) it offers a large-scale, coarse-grained view of an organism at the mRNA level b) gives important indications for complex diseases c) assist in the development of target and personalized medicines d) helps in understanding evolution by comparing genetic networks of various genomes and e) explains how different phenotypes emanate and which groups of genes are responsible for them.

Pure EAs are useful tool to analyze only small networks. However, it can be hybridized to enhance its efficiency for large networks. Although hybridized methods are computationally expensive and perform well with small-size networks but they become less efficient when analyzing the large-size networks. Scalability and parallelism can be achieved at several-level, from individual evaluation to iterative and breaking the entire problem into sub-problems. Various methods such as PSOs, RNNs, FLs, ANNs, ODEs etc. has been fused with EAs to get some good results. One of the important issues with the inference of GRN from microarray data is both limited and noisy nature of these data. It shows the compelling need to look for time-series data from various sources and some other types of biological data such as ChIP, knockout microarray experiments, protein-protein interactions and miRNA interference data. These types of data can be included in EAs in a multi-objective setting in order to speed up convergence.